\documentclass{ws-procs9x6-cpt19}
\begin{document}

\newcommand{\refeq}[1]{(\ref{#1})}
\def\etal {{\it et al.}}
\def\cf {{\it cf.}}

\title{Constraints on Lorentz Invariance Violations from Gravitational Wave Observations}

\author{Anuradha Samajdar$^1$}

\address{$^1$Nikhef,\\
Amsterdam, 105 Science Park, 1098 XG, The Netherlands}

\author{for the LIGO Scientific Collaboration and Virgo Collaboration}

\begin{abstract}
Using a deformed dispersion relation for gravitational waves, Advanced LIGO 
and Advanced Virgo have been able to place constraints on violations of local Lorentz 
invariance as well as the mass of the graviton. We summarise the method to obtain the current bounds from the 10 
significant binary black hole detections made during the first and second
observing runs of the above detectors. 
\end{abstract}

\bodymatter

\section{Introduction}\label{sec:intro}
The year 2015 saw the advent of gravitational wave (GW) astronomy 
with GW150914~\cite{gw150914}, the first directly detected GW signal 
from a binary black hole (BBH) merger.
Ref.\ \refcite{tgr_companion_150914} performed
tests on strong-field gravity in the highly dynamical 
regime of general relativity (GR), finding no 
statistically significant violations of GR.
Since then, 10 significant BBH signals have been detected, in 
addition to a binary neutron star (BNS) signal~\cite{o2catalogue}.
The first constraints on local Lorentz invariance 
violation (LIV) using real GW data were reported in Ref.\ \refcite{gw170104}. 
These bounds have been revised recently 
and reported in Ref.\ \refcite{tgro2}. These bounds, 
however, rely on the propagation effects and therefore do not directly probe the 
dynamical regime of gravity.

In this proceedings, we give a brief overview of the method to constrain LIV in Sec.~\ref{sec:method} and
summarise the results with some concluding remarks in Sec.~\ref{sec:res}.

\section{Method}
\label{sec:method}
GWs propagating in GR are non-dispersive and travel with the speed 
of light. Following Refs.\ \refcite{MYW,YYP}, we adopt the generic 
dispersion relation
\begin{equation}
E^2 = p^2c^2 + A_\alpha p^\alpha c^\alpha.
 \label{eq:mdr}
\end{equation}
This is a Lorentz violating dispersion relation for $\alpha>0$, 
the LIV parameter is characterised by $A_\alpha$. 
$\alpha=0$ is a special case where 
we may parameterise the additional term in Eqn.~\ref{eq:mdr} as $A_0=m_g^2 c^4$, 
$m_g$ being the mass of the graviton. 
Examples of Lorentz violating theories for specific forms of Eqn.~\ref{eq:mdr} include 
Doubly Special Relativity~\cite{dsr} for $\alpha=3$ and 
Ho{\v r}ava-Lifshitz theory~\cite{hlt} for $\alpha=4$, \cf\ Refs.\ \refcite{gw170104}, \refcite{tgro2} 
for more examples and corresponding references. As noted in Ref.\ \refcite{gw170104}, a combination of values 
of $\alpha$ and the sign of $A_\alpha$ can indicate whether the speed of GWs is  
subluminal or superluminal.

In the presence of dispersion, the low (high)-frequency components of a GW signal
travel slower (faster) and result in an overall offset in arrival times at the detector, 
leading to a frequency-dependent shift in the phasing. 
In frequency domain (FD), the total phase is then given by $\Psi(f) = \Psi_{GR}(f) + \Psi_\alpha(f)$. 
$\Psi_{GR}(f)$ is the phasing obtained from GR predictions and 
$\Psi_\alpha(f)$ denotes the phase shift following from the 
dispersion. 
The waveform model in FD used in our analyses is constructed by 
$\tilde{h}(f) = \mathcal{A}(f)\mathrm{e^{-i\Psi}}$.
We associate a length-scale $\lambda_A = hc|A_\alpha|^{1/(\alpha-2)}$ with the 
LIV parameter, 
where $h$ is the Planck's constant and $c$ is the speed of light. 
$\lambda_A$ may be thought of as a screening length
corresponding to an effective gravitational potential. In terms of 
$\lambda_A$, the phasing relations are given by
\begin{equation}
\Psi_\alpha(f) =
\begin{cases}
  \mathrm{sign}(A_\alpha)\frac{\pi D_1}{|\lambda_A|}\ln(\pi \mathcal{M} f), \ \text{if } \alpha=1,\\
 -\mathrm{sign}(A_\alpha) \frac{\pi}{(1-\alpha)}\frac{D_\alpha}{|\lambda_A|^{2-\alpha}} \frac{f^{\alpha-1}}{(1+Z)^{1-\alpha}}, \ \text{if } \alpha \ne 1. 
\label{eq:dpsi-liv}
 \end{cases} 
\end{equation}
In the above equation, $\mathcal{M}$ is the detector-frame chirp mass of the binary system, a combination 
of component masses given by $\mathcal{M}=(m_1m_2)^{3/5}/(m_1+m_2)^{1/5}$, $m_1$ and $m_2$ being the component 
masses. $f$ is the frequency component and $Z$ denotes the redshift to the source. 
$D_\alpha$ is a cosmological distance, see Refs.\ \refcite{MYW}, \refcite{tgro2} for more details.

The analyses carried out in the following section is based on a Bayesian framework which 
incorporates the Bayes' theorem $p(\vec{\theta}|d) = p(d|\vec{\theta}) p(\vec{\theta})/p(d)$, where 
$\vec{\theta}$ refers to a parameter set, $d$ refers to the data, $p(\vec{\theta}|d)$ refers to 
the posterior probability density obtained on $\vec{\theta}$ from the likelihood calculated from the data 
$p(d|\vec{\theta})$ and the a priori probability density given by $p(\vec{\theta})$. $p(d)$ is a 
normalisation constant. The information learnt from the data is folded in the likelihood which takes the 
following form
\begin{equation}
 p(d|\vec{\theta}) \propto \exp{\left [-\frac{1}{2}(d-h|d-h)\right ]}.
 \label{eqn:lhood}
\end{equation}
In the presence of a GW signal, the data output from the detector is $d = h(t) + n(t)$,
where $h(t)$ is the GW signal and $n(t)$ is the noise. 
For our analyses, the likelihood integral is computed in FD 
by including the LIV-deformed phase in the model waveform. 
For a value of $\alpha$, this enables us to obtain a posterior probability density 
function on the parameter $A_\alpha$, leading to a constraint on LIV.

\section{Results}
\label{sec:res}
\begin{figure}
\begin{center}
\includegraphics[width=4in]{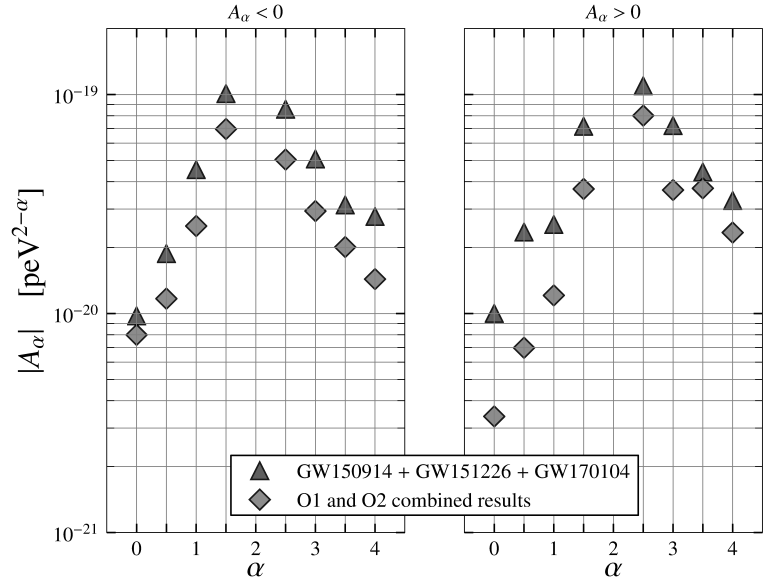}
\end{center}
\caption{90\% credible upper bounds on $A_\alpha$ from the BBH detections GW150914, GW151226 and GW170104 (triangles) and those from 
         combining all 10 significant BBH detections (diamonds) in O1 and O2 as given in Ref. 7.
         Same as Fig.5 of Ref. 7 but grayscaled.}
\label{fig1}
\end{figure}

Being a propagation effect, the strongest constraints come from 
events located at larger luminosity distances. 
The bounds obtained from the catalogue of 10 sources are presented in Fig.~\ref{fig1}.
The current bounds obtained from combining all sources lead to an improvement in previously reported bounds~\cite{gw170104} 
by factors up to 2.4 as reported in Ref.\ \refcite{tgro2}.

From combining these sources, the mass of the graviton has 
been constrained to $m_g \le 5.0\times10^{-23} \ \mathrm{eV/c^2}$ at 90\% confidence. 

\section*{Acknowledgments}
The authors gratefully acknowledge the support of the United States
National Science Foundation (NSF) for the construction and operation of the
LIGO Laboratory and Advanced LIGO as well as the Science and Technology Facilities Council (STFC) of the
United Kingdom, the Max-Planck-Society (MPS), and the State of
Niedersachsen/Germany for support of the construction of Advanced LIGO 
and construction and operation of the GEO600 detector. 
Additional support for Advanced LIGO was provided by the Australian Research Council.
The authors gratefully acknowledge the Italian Istituto Nazionale di Fisica Nucleare (INFN),  
the French Centre National de la Recherche Scientifique (CNRS) and
the Foundation for Fundamental Research on Matter supported by the Netherlands Organisation for Scientific Research, 
for the construction and operation of the Virgo detector
and the creation and support  of the EGO consortium. 
The authors also gratefully acknowledge research support from these agencies as well as by 
the Council of Scientific and Industrial Research of India, 
the Department of Science and Technology, India,
the Science \& Engineering Research Board (SERB), India,
the Ministry of Human Resource Development, India,
the Spanish  Agencia Estatal de Investigaci\'on,
the Vicepresid\`encia i Conselleria d'Innovaci\'o, Recerca i Turisme and the Conselleria d'Educaci\'o i Universitat del Govern de les Illes Balears,
the Conselleria d'Educaci\'o, Investigaci\'o, Cultura i Esport de la Generalitat Valenciana,
the National Science Centre of Poland,
the Swiss National Science Foundation (SNSF),
the Russian Foundation for Basic Research, 
the Russian Science Foundation,
the European Commission,
the European Regional Development Funds (ERDF),
the Royal Society, 
the Scottish Funding Council, 
the Scottish Universities Physics Alliance, 
the Hungarian Scientific Research Fund (OTKA),
the Lyon Institute of Origins (LIO),
the Paris \^{I}le-de-France Region, 
the National Research, Development and Innovation Office Hungary (NKFIH), 
the National Research Foundation of Korea,
Industry Canada and the Province of Ontario through the Ministry of Economic Development and Innovation, 
the Natural Science and Engineering Research Council Canada,
the Canadian Institute for Advanced Research,
the Brazilian Ministry of Science, Technology, Innovations, and Communications,
the International Center for Theoretical Physics South American Institute for Fundamental Research (ICTP-SAIFR), 
the Research Grants Council of Hong Kong,
the National Natural Science Foundation of China (NSFC),
the Leverhulme Trust, 
the Research Corporation, 
the Ministry of Science and Technology (MOST), Taiwan
and
the Kavli Foundation.
The authors gratefully acknowledge the support of the NSF, STFC, MPS, INFN, CNRS and the
State of Niedersachsen/Germany for provision of computational resources.

\end{document}